\documentclass[a4paper,oneside]{article}

\usepackage{graphicx}
\usepackage{subfigure}
\usepackage{cite}
\usepackage{anysize}
\pdfoutput=1
\DeclareGraphicsExtensions{.jpg}

\begin{document}

\vskip 1cm
\marginsize{3cm}{3cm}{3cm}{1cm}

\begin{center}
{\bf{\huge Investigation of Ion Backflow in Bulk Micromegas Detectors}}\\
~\\
Purba Bhattacharya$^a$, Deb Sankar Bhattacharya$^{b,c}$, Supratik Mukhopadhyay$^b$, Sudeb Bhattacharya$^d$, Nayana Majumdar$^b$, Sandip Sarkar$^b$, Paul Colas$^c$, David Attie$^c$\\
{\em $^a$ School of Physical Sciences, National Institute of Science Education and Research, Jatni, Bhubaneswar - 752005, India}\\
{\em $^b$ Applied Nuclear Physics Division, Saha Institute of Nuclear Physics, Kolkata - 700064, India}\\
{\em $^c$ DSM/IRFU, CEA/Saclay, F-91191 Gif-sur-Yvette CEDEX, France}\\
{\em $^d$ Retired Senior Professor, Applied Nuclear Physics Division, Saha Institute of Nuclear Physics, Kolkata - 700064, India}\\
~\\
~\\
~\\
~\\
~\\
{\bf{\large Abstract}}
\end{center}

The operation of gas detectors is often limited by secondary effects, originating from avalanche-induced photons and ions.
Ion backflow is one of the effects limiting the operation of a gas detector at high flux, by giving rise to space charge which disturbs the electric field locally.
For the Micromegas detector, a large fraction of the secondary positive ions created in the avalanche can be stopped at the micro-mesh.
The present work involves  measurements of the ion backflow fraction (using an experimental setup comprising of two drift planes) in bulk Micromegas detectors 
as a function of detector design parameters. These measured characteristics have also been compared in detail to 
numerical simulations using the Garfield framework that combines packages such as neBEM, Magboltz and Heed.
Further, the effect of using a second micro-mesh on ion backflow and other parameters has been studied numerically.

\vskip 1.5cm
\begin{flushleft}

\end{flushleft}

\vskip 1.5in
\noindent
{\bf ~$^*$Corresponding Author}: Purba Bhattacharya

E-mail: purba.bhattacharya85@gmail.com

\newpage

\section{Introduction}
\label{sec:intro}

Micro Pattern Gaseous Detectors (MPGDs) \cite{MPGD} have played an important role in radiation detection and 
imaging in many fields such as particle-physics and nuclear-physics experiments, astro-particle research, medical 
imaging, material science etc \cite{TPC1}.
Despite their widespread use, the operation of gas detectors has often suffered from some secondary effects, 
originating from avalanche-induced photons and ions. One of these secondary effects is ion backflow which is the 
drift of the positive ions produced in the avalanche, from the amplification region towards the drift volume.
In a high-rate experiment, it is very important to limit ion backflow from the amplification region, in order 
to avoid the distortion of the local electric field in the drift volume which may affect the drift behaviour of 
the electrons from a later track.
The backflow fraction of Micro Pattern Gaseous Detectors is smaller than that of the more classical 
configuration. For example, in the Micromegas detector, a large fraction of the secondary positive ions created 
in the avalanche can be stopped at the micro-mesh depending on the field ratio, detector geometry, etc \cite{IBFPaul}.

\begin{figure}[hbt]
\centering
\includegraphics[scale=0.25]{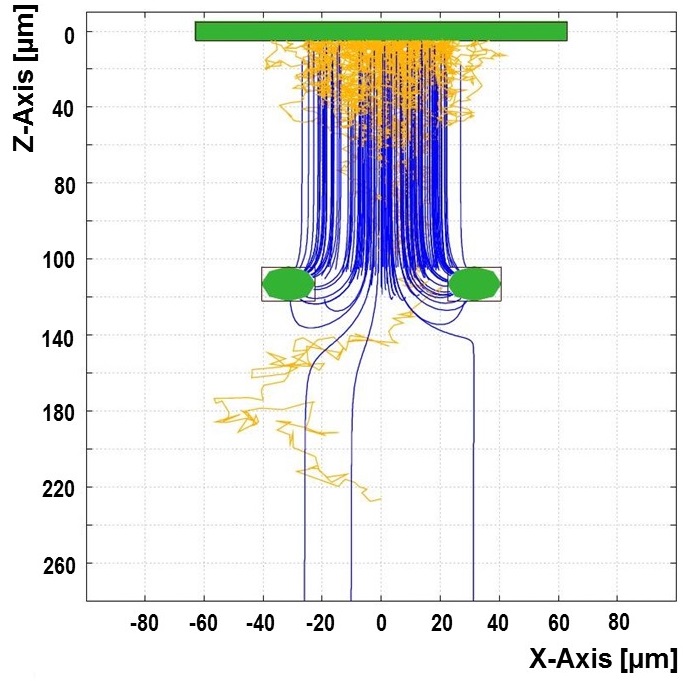}
\caption{Electron avalanche and ion drift lines.}
\label{IBF-1}
\end{figure}

The Micromegas (MICRO-MEsh GAseous Structure) \cite{Micromegas}, is a parallel plate device and composed of a very thin metallic micro-mesh, which separates the low-field drift region from the high-field amplification region.
Due to the field gradient between the drift and amplification regions and the periodic hole pattern, the field lines from the drift region are compressed in the vicinity of the micro-mesh holes and form a funnel having width of a few microns in the amplification region.
As a result, an electron approaching the micro-mesh is focused towards the center of a hole and produces an avalanche inside the funnel.
Due to the transverse diffusion, the avalanche also extends outside the funnel.
Conversely, the ions, due to their larger mass, are not affected much by the diffusion and drift along the field lines.
Most of the ions, created outside the funnel, follow the field lines and are collected by the micro-mesh \cite{Chefdeville}.
A very small fraction, produced in the thin funnel, drifts back towards the drift volume (Figure \ref{IBF-1}).

The ion backflow fraction can, thus, be defined as:
\begin{eqnarray}
IBF = \frac {N_b} {N_t}
\end{eqnarray}

\noindent where $N_t$  is the average number of ions produced in an electron avalanche and $N_b$ the average number of the backflowing ions. 
When the transverse spread of the avalanche is small w.r.t the hole pitch, in the two dimensional limit (neglecting the longitudinal development) it can been shown that \cite{Chefdeville},

\begin{eqnarray} 
\label{ibfeqn}
IBF \propto \frac {1} {FR} \bigg(\frac {p} {\sigma_t}\bigg)^2 
\end{eqnarray}

\noindent Here $FR$ is the field ratio (amplification field ($E_{amp}$)/drift field ($E_{drift}$)) and $p$ is the mesh pitch.
$\sigma_t$ is related to the transverse diffusion of the electron and given by $D_t\sqrt{z}$ wherein $D_t$ is the transverse diffusion coefficient of electron and $z$ is the path traversed.
Thus, the backflow fraction can be reduced by optimizing the detector geometry and gas parameters.

The bulk Micromegas detector \cite{Bulk} has been considered to be one of the good choices for a read-out system in various experiments due to its performances in terms of gain uniformity, energy and space point resolutions and low ion-feedback \cite{TPC2, BULK128, TPC3, TPC4}.              
In this work, experimental and numerical studies have been carried out to illustrate the effects of different amplifications gaps and mesh hole pitches on the ion backflow fraction for bulk Micromegas detectors in $\mathrm{Argon}$-$\mathrm{Isobutane}$ ($90:10$) gas mixture.
In conjunction with the experimental work, extensive numerical simulations have been carried out using Garfield to enrich our understanding of the complex physical processes occurring within the device.
Finally, the effects of using a second micro-mesh \cite{IBF2} on various parameters, including ion backflow, have been studied numerically.

\section{Experimental Setup}
\label{sec:experiment}

Several small size detectors with an active area of $15~\mathrm{{cm}^2}$ have been fabricated at CEA, Saclay, France and tested at SINP, Kolkata, India.
Each of the bulk Micromegas detectors is equipped with a calendered woven micro-mesh.
The design parameters of the detectors are compiled in Table \ref{design}.

\begin{table}[hbt]
\caption{Design parameters of the bulk Micromegas detectors. All detectors have wire diameter of $18~\mu\mathrm{m}$.}\label{design}
\begin{center}
\begin{tabular}{|c|c|}
\hline
Amplification Gap (in $\mu\mathrm{m}$) & Mesh Hole Pitch (in $\mu\mathrm{m}$)\\
\hline
64 & 63 \\
\hline
128 & 63 \\
\hline
128 & 78 \\
\hline
192 & 63 \\
\hline
\end{tabular}
\end{center}
\end{table}

\begin{figure}[hbt]
\centering
\includegraphics[scale=0.5]{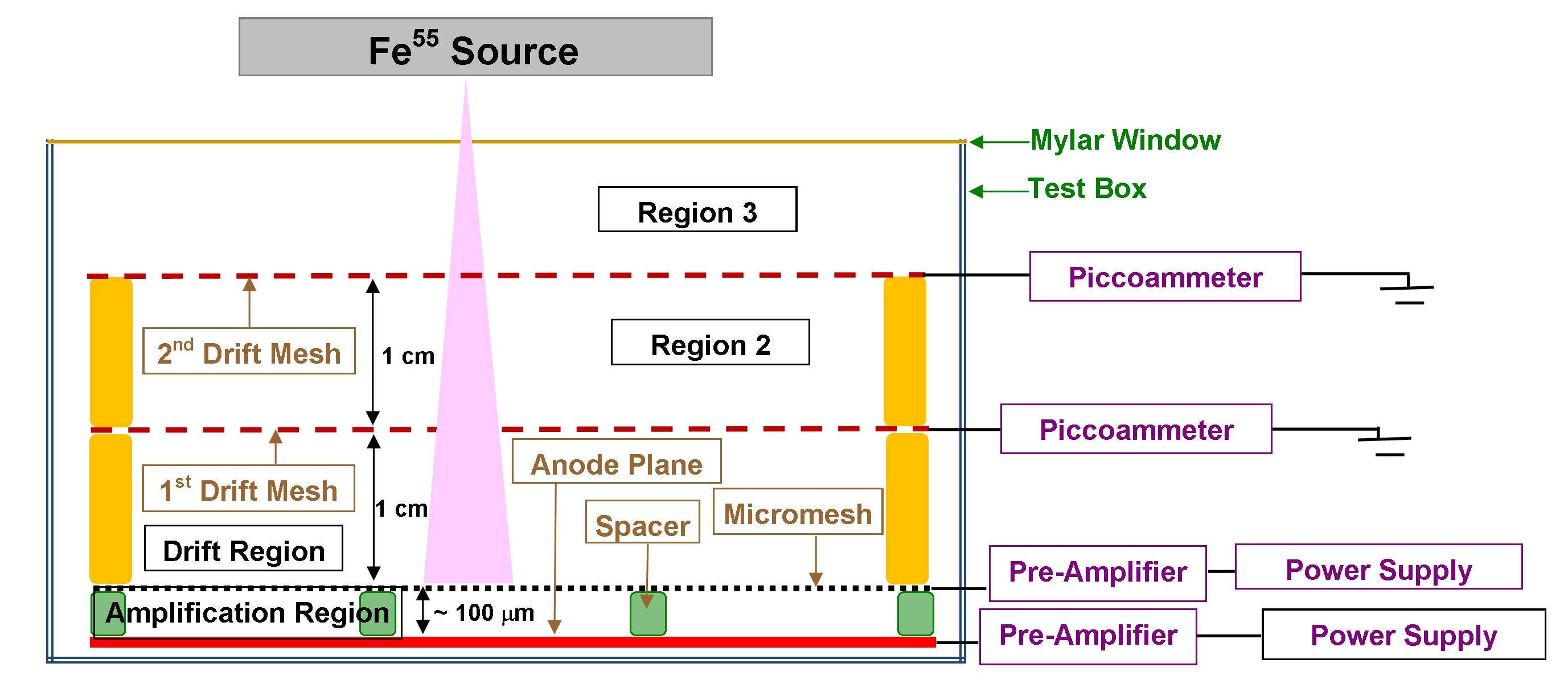}
\caption{Schematic diagram of experimental setup for ion backflow measurement.}
\label{ExptSetup}
\end{figure}

At SINP, an experimental setup (Figure \ref{ExptSetup}) has been built to measure the ion backflow.
The detector, as shown in the figure, is placed inside a leak-proof chamber and the drift plane is  mounted $1~\mathrm{cm}$ above the detector.
A $^{55}\mathrm{Fe}$ source has been used to produce the primary electrons in the drift region.
In the test box, a provision has been made to move and position the source at a desired location within the chamber, without breaking the gas flow, by using a Wilson seal .

\begin{figure}[hbt]
\centering
\subfigure[]
{\label{Testbox-1}\includegraphics[scale=0.055]{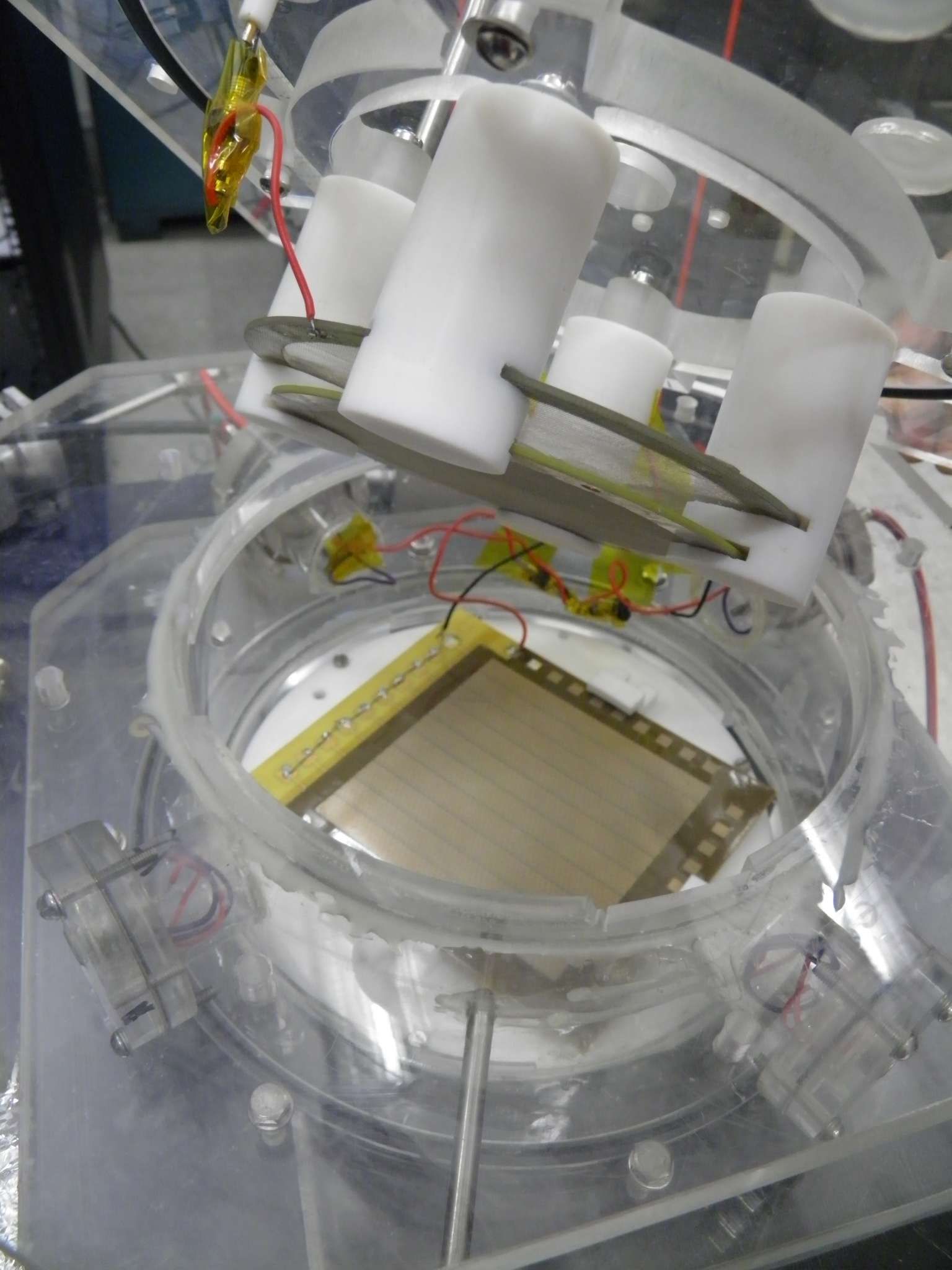}}
\subfigure[]
{\label{Testbox-2}\includegraphics[scale=0.081]{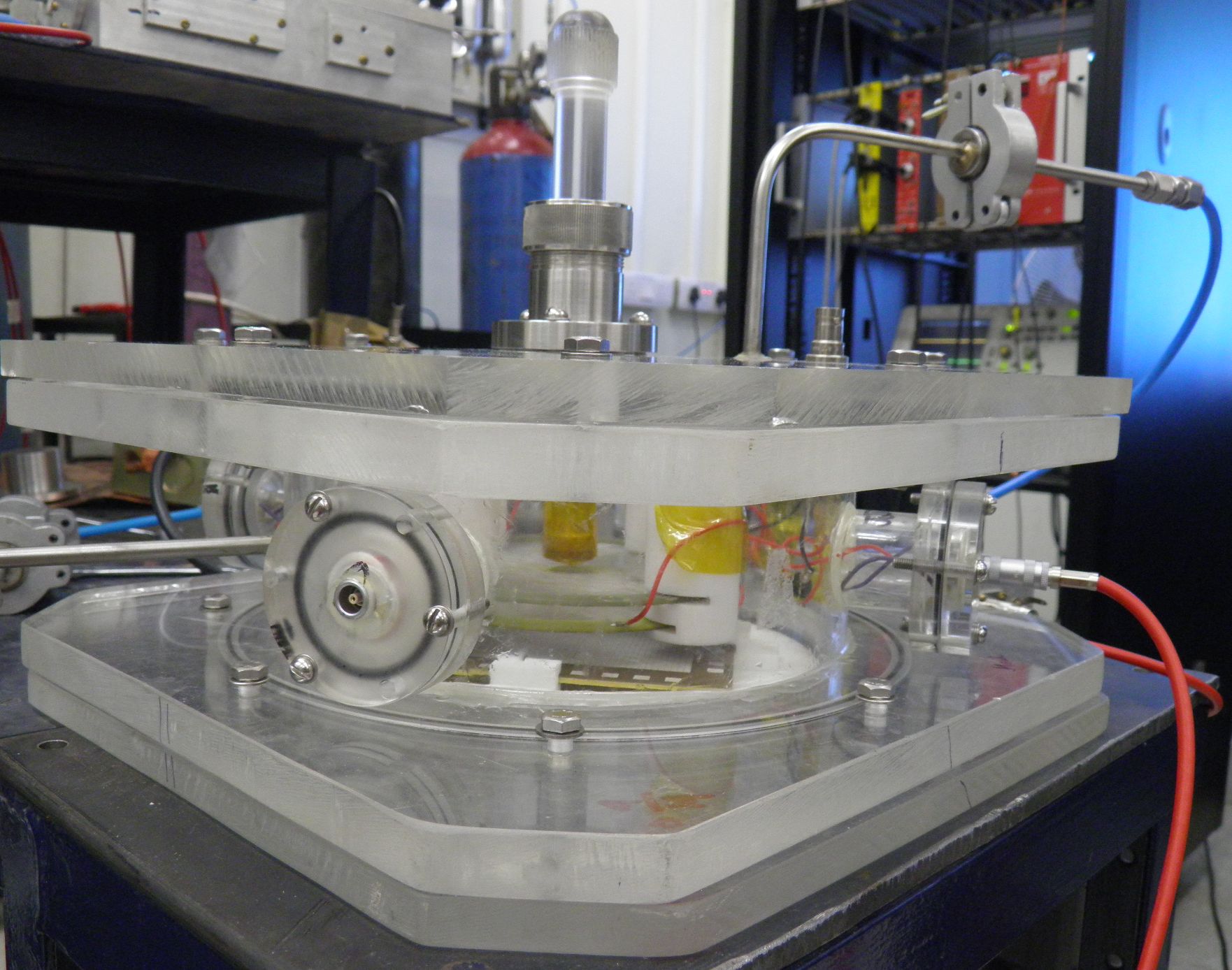}}
\caption{Photograph of (a) the test box with the detector and two drift mesh, (b) the whole setup.}
\label{Testbox}
\end{figure}

As already discussed in our earlier work \cite{IBF-Purba}, the use of a single drift mesh is likely to generate additional contributions to the estimates of ion backflow.
Along with the contribution of the ions from the avalanche and the drift region, there is a possibility of having an additional contribution to the drift current from the ions created in Region 2 in Figure \ref{ExptSetup}. 
So the setup has been modified by placing a second drift mesh at a distance of $1~\mathrm{cm}$ above the first one and this has been  kept at the same voltage as that of the first drift mesh (Figure \ref{ExptSetup}).
The ions that are created between the test box window and the upper drift mesh (Region 3 in Figure \ref{ExptSetup}) are collected on the outer drift mesh ($2^{nd}$ Drift Mesh in Figure \ref{ExptSetup}).
Thus, the current on the inner drift mesh ($1^{st}$ Drift Mesh in Figure \ref{ExptSetup}) can now be considered as a better estimate of the ionic current from the avalanche.
The photograph of the test box using two drift meshes and the positioning of the source are shown in Figure \ref{Testbox}.

The currents on the drift mesh and the micro-mesh have been measured and the ratio of these two currents gives an estimate of the backflow fraction.
\begin{eqnarray}
IBF =  \frac {I_C} {(I_M + I_C)}
\end{eqnarray}

\noindent where $I_C$ is the current measured on the $1^{st}$ drift mesh (Figure \ref{ExptSetup}) and is proportional to the number of ions collected on the drift mesh;
$I_M$ is the current measured in the micro-mesh and proportional to the number of ions collected on the mesh.
For the measurement of current, a pico-ammeter (CAEN model AH401D) has been used (Figure \ref{ExptSetup}) which can measure the current only from an electrode which is at a ground potential.
Because of this, the potential configuration, in the present experiment, has been suitably altered depending on the nature of measurement.
For example, in one of the configurations, the detector has been biased with a negative voltage at the micro-mesh plane, more negative voltage at the drift plane and the anode connected to the ground.
In another configuration, for the measurement of current from the mesh, the micro-mesh has been grounded, whereas the anode plane and the drift plane have been biased with more positive and more negative
voltages, respectively, with respect to the mesh plane, maintaining the proper field configuration.
Similarly, for measuring $I_C$, the drift plane is set to the ground potential and the mesh and the anode to the required positive high voltages.
For the measurement of the anode current, a current integrator (Danfysik 554) has been used.
The positive currents from the micro-mesh and the drift mesh have also been cross-verified using the current integrator.

\section{Simulation Tools}
The experimental data have been compared with the estimates obtained through numerical simulation.
We have used the Garfield \cite{Garfield1, Garfield2} simulation framework.
This framework was augmented in 2009 through the addition of the neBEM (nearly exact Boundary Element Method \cite{neBEM1, neBEM2, neBEM3, neBEM4, neBEM5}, which is known to be very accurate throughout the computational domain) toolkit to carry out 3D electrostatic field simulation.
Besides the neBEM, the Garfield framework provides interfaces to the programs HEED \cite{HEED1, HEED2} for the primary ionization calculation and Magboltz \cite{Magboltz1, Magboltz2} for computing the drift, diffusion, Townsend and attachment coefficients.

In the numerical simulation, the electrons have been injected on square areas with the sides equal to that of the size of the pitch.
The first area has been placed $100~\mu\mathrm{m}$ above the mesh and the subsequent areas have been placed from $1~\mathrm{mm}$ to $1.0~\mathrm{cm}$ with a spacing of $1~\mathrm{mm}$ between them. 
Each square is subdivided into $10\times10$ grid pattern. 
Each node on the square has been injected with an electron 100 times.
As a result, $100\times100\times11$ electrons are made to drift towards the amplification region where they have a chance to multiply.
The primary ions in the drift region and the ions created in the avalanche have been considered for the estimation.
The backflow fraction has been calculated as

\begin{eqnarray}
IBF = \frac {N_{id}} {(N_{id}+N_{im})}
\end{eqnarray}

\noindent where ${{N}_{id}}$ is the number of ions collected at the drift plane and ${{N}_{im}}$ is the number of ions collected at the micro-mesh.

\section{Results}

\begin{figure}[hbt]
\centering
\subfigure[]
{\label{IBF-III-1}\includegraphics[scale=0.05]{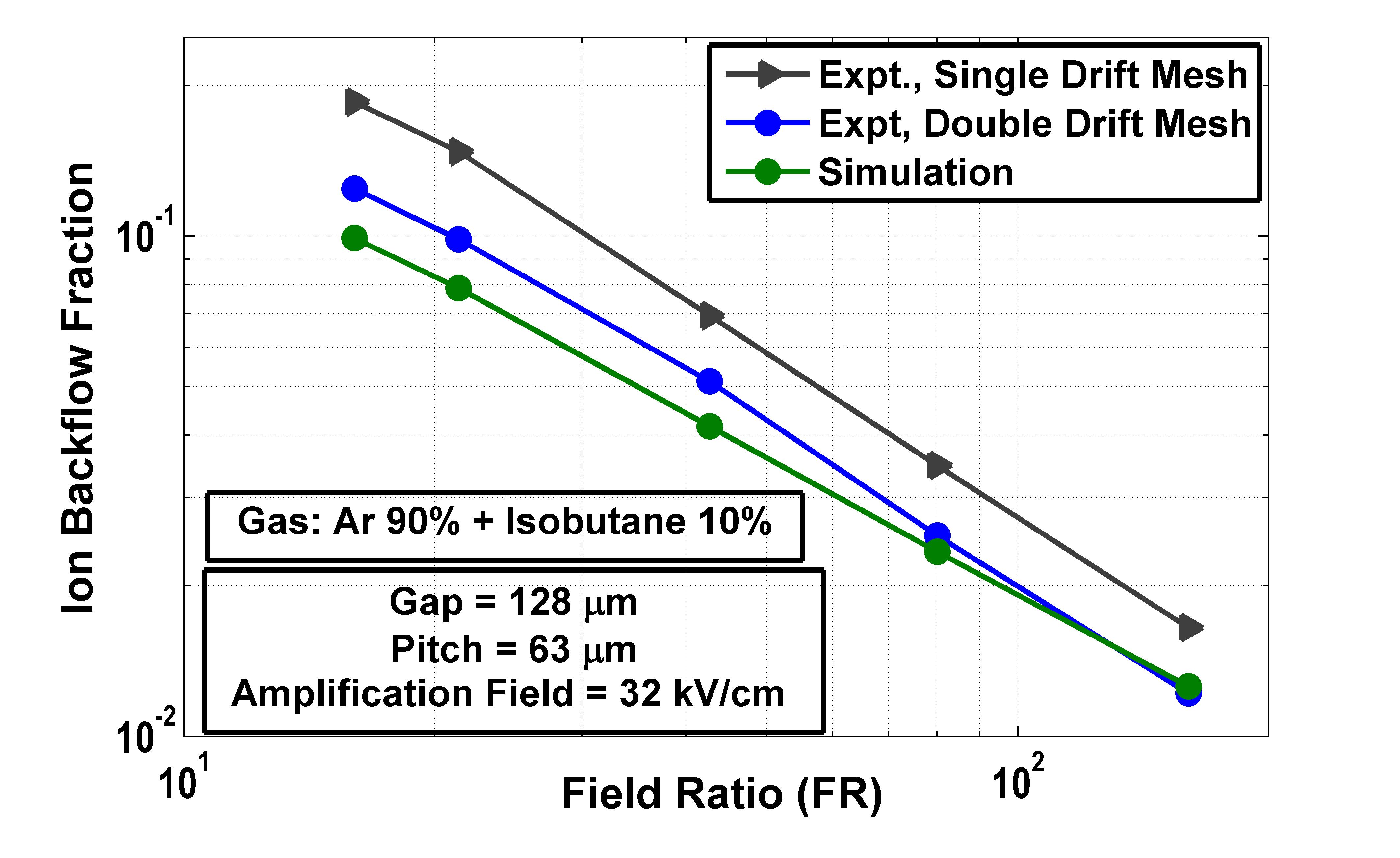}}
\subfigure[]
{\label{IBF-III-2}\includegraphics[scale=0.05]{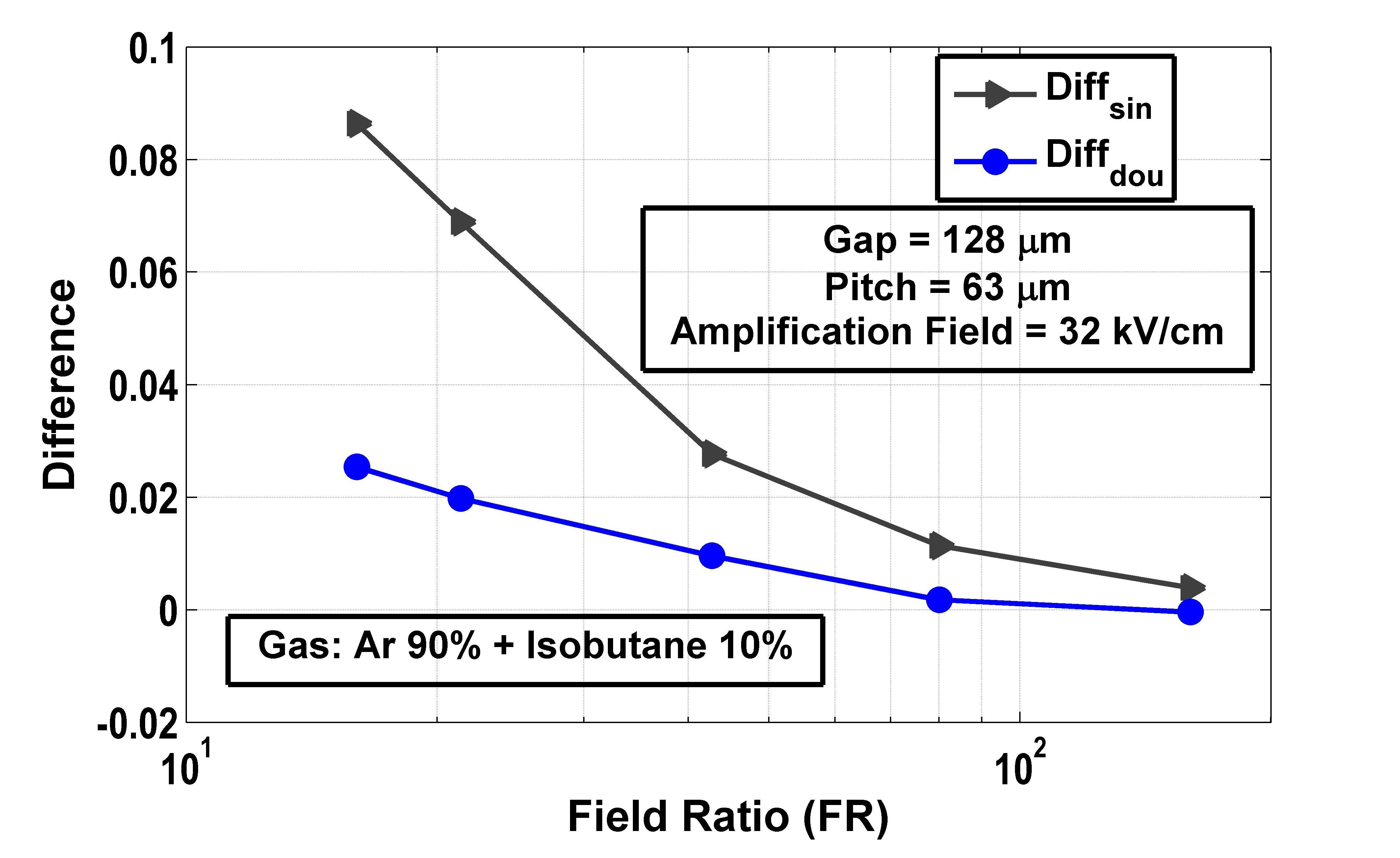}}
\subfigure[]
{\label{IBF-III-3}\includegraphics[scale=0.07]{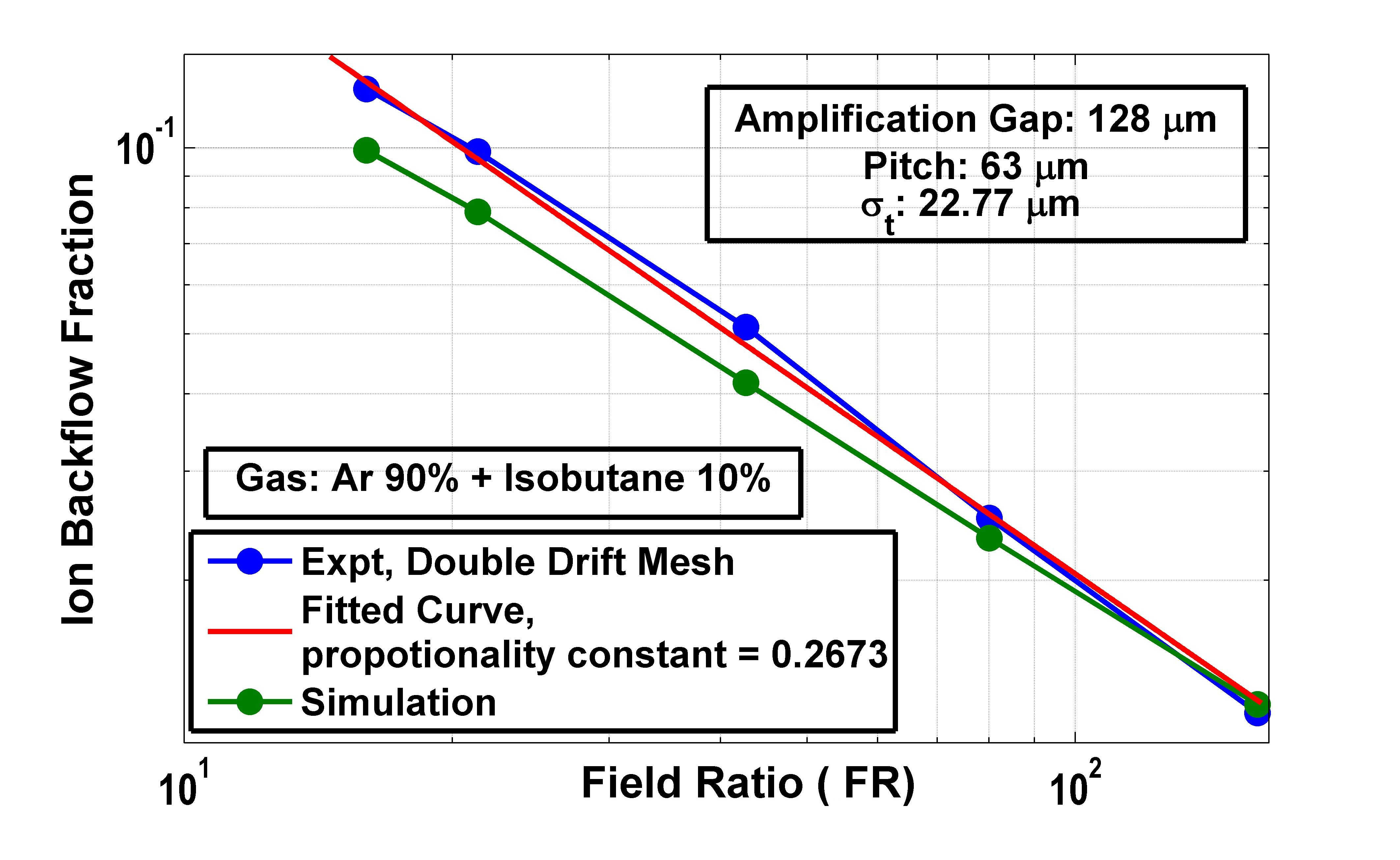}}
\caption{Variation of (a) the ion backflow fraction with the field ratio ($FR$) for a particular amplification field; (b) the difference between the simulated results and the experimental data; (c)  comparison with fitted curve using experimental data point and Eqn. 1.2 in \(\mathrm{Argon}\)-\(\mathrm{Isobutane}\) (\(90:10\)) mixture.}
\label{IBF-III}
\end{figure}

The variation of the measured ion backflow fraction with the field ratio ($FR$) is shown in Figure \ref{IBF-III-1} for the $128~\mu\mathrm{m}$ bulk Micromegas detector (pitch $63~\mu\mathrm{m}$).
For this measurement, the amplification field has been kept constant while the drift field has been varied.
The trend is similar for both single and double drift mesh setups.
The difference between the simulated results with the experimental data has been estimated and plotted in Figure \ref{IBF-III-2}.
${Diff_{sin}}$ represents the difference between the simulated data and the single mesh measurements while ${Diff_{dou}}$ takes care of the double mesh measurements.
As shown in Figure \ref{IBF-III-2} the numerical estimates are closer to the experimental data in the double drift mesh case. Thus the numerical estimates in this case are expected to provide  better estimates of the backflow fraction.
The experimental data points (for the double drift mesh) have been also fitted using Eqn. \ref{ibfeqn} as shown in Figure \ref{IBF-III-3}.
The three curves agree quite well with each other.
It may be noted here that, while carrying out the fit, the value of $\sigma_t$ corresponding to the amplification field has been used. This is because of the fact that the transverse diffusion in that region is expected to determine the $IBF$.

\begin{figure}[hbt]
\centering
\subfigure[]
{\label{IBF-IV-1}\includegraphics[scale=0.05]{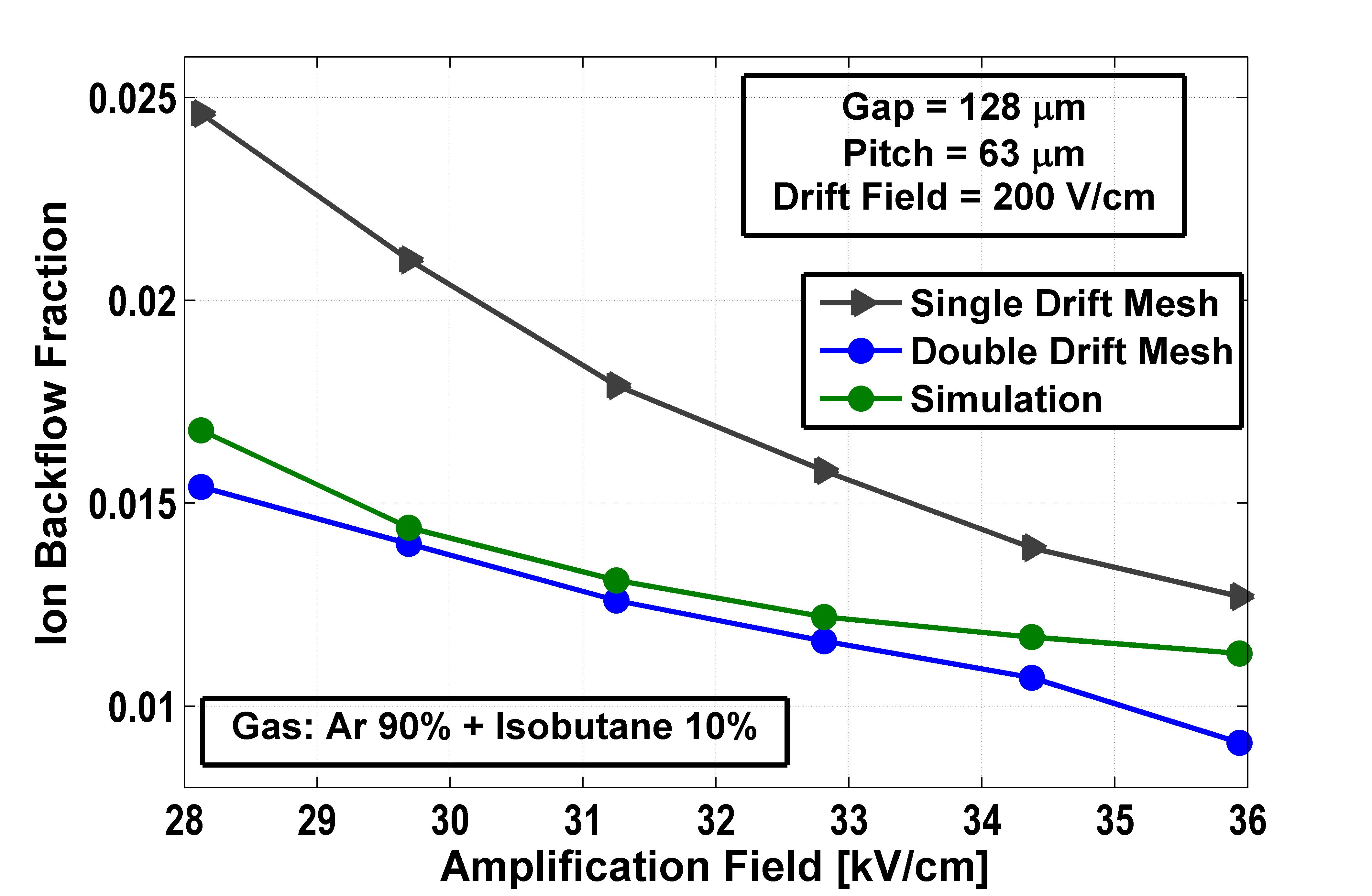}}
\subfigure[]
{\label{IBF-IV-2}\includegraphics[scale=0.05]{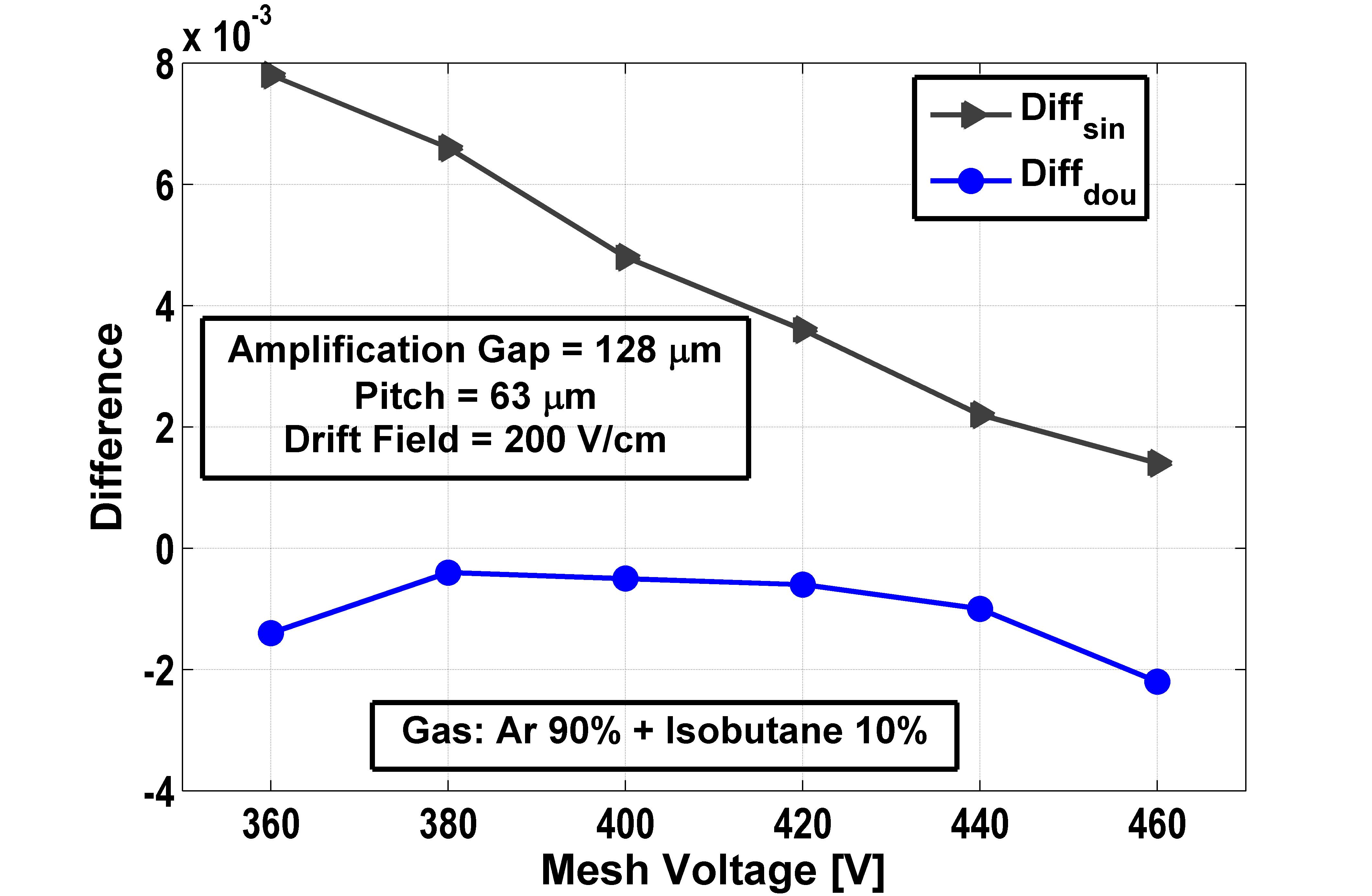}}
\caption{Variation of (a) the ion backflow fraction with the amplification field for a particular drift field of $200~\mathrm{V/cm}$; (b) the difference between the simulated results and the experimental data in $\mathrm{Argon}$-$\mathrm{Isobutane}$ (${90:10}$) mixture.}
\label{IBF-IV}
\end{figure}

For a particular detector, at a fixed drift field, the increase of the amplification field reduces the backflow fraction due to the combined effect of the transverse extent of the avalanche spread and field ratio, as shown in Figure \ref{IBF-IV-1}.
Here also, the difference between the simulated estimates and the experimental data using the double drift mesh is less (Figure \ref{IBF-IV-2})compared to that in the case of single drift mesh.

\subsection{Effect of Detector Geometry}

\begin{figure}[hbt]
\centering
\subfigure[]
{\label{IBF-V-1}\includegraphics[scale=0.05]{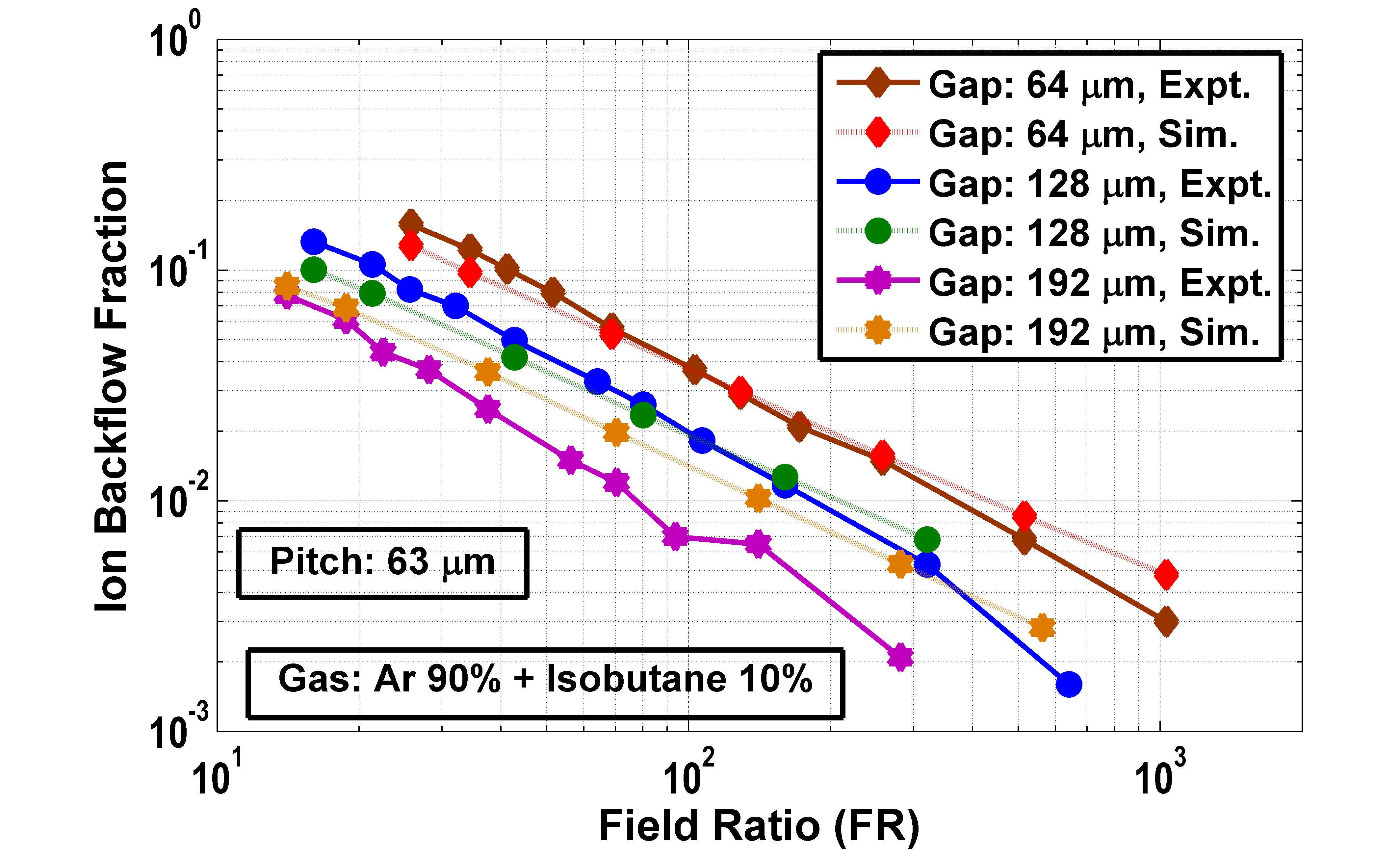}}
\subfigure[]
{\label{IBF-V-2}\includegraphics[scale=0.05]{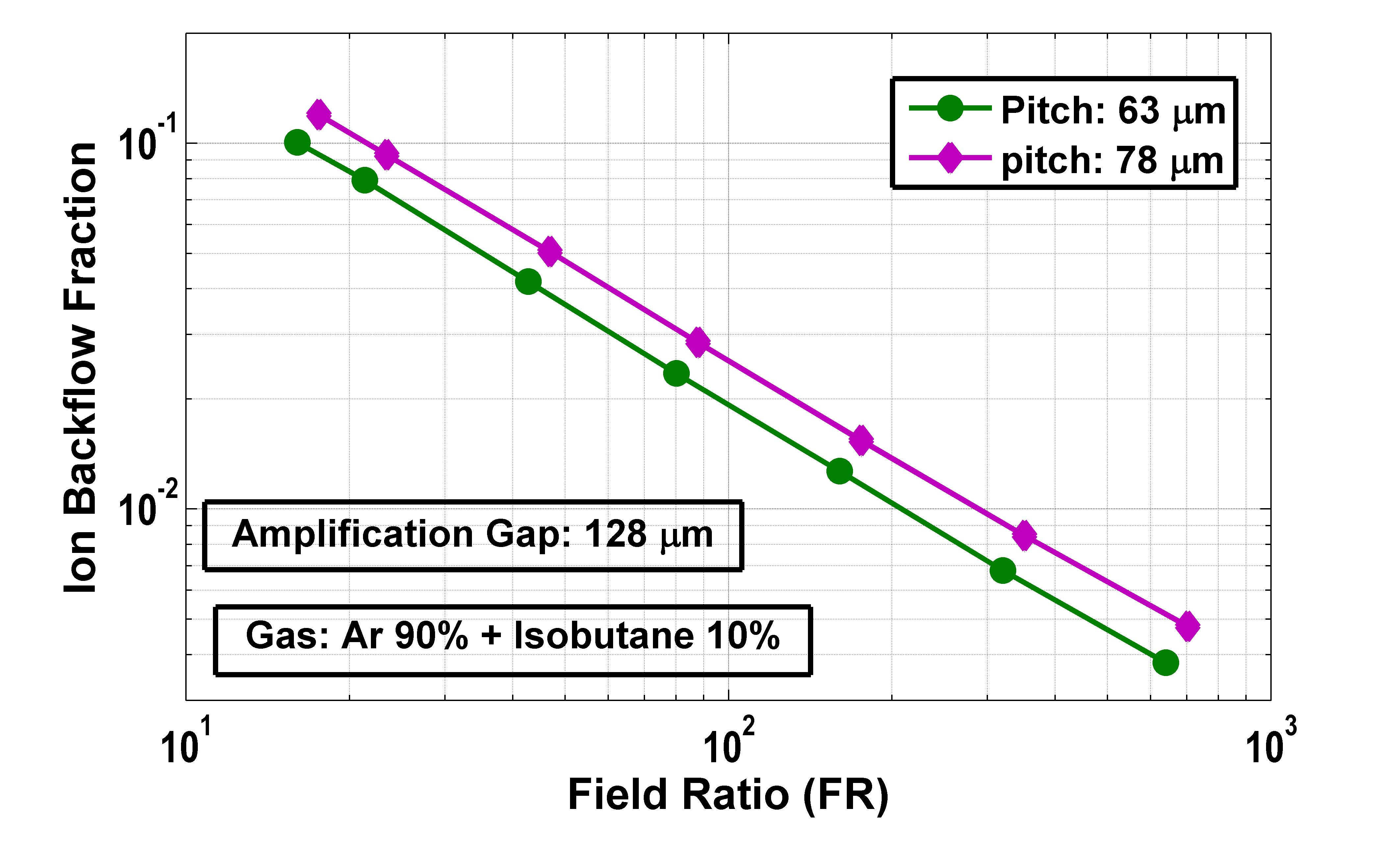}}
\caption{In $\mathrm{Argon}$-$\mathrm{Isobutane}$ (${90:10}$) mixture, variation of the ion backflow fraction with the field ratio ($FR$) for (a) three different amplification gap having same mesh hole pitch [gap = $64~\mu\mathrm{m}$, pitch = $63~\mu\mathrm{m}$, mesh voltage = $-330~\mathrm{V}$, gain = $\sim3000$; gap = $128~\mu\mathrm{m}$, pitch = $63~\mu\mathrm{m}$, mesh voltage = $-410~\mathrm{V}$, gain $\sim3200$; gap = $192~\mu\mathrm{m}$, pitch = $63~\mu\mathrm{m}$, mesh voltage = $-540~\mathrm{V}$, gain $\sim4000$]; (b) two different mesh hole pitch having same amplification gap [gap = $128~\mu\mathrm{m}$, pitch = $63~\mu\mathrm{m}$, mesh voltage = $-410~\mathrm{V}$, gain $\sim3200$; gap = $128~\mu\mathrm{m}$, pitch = $78~\mu\mathrm{m}$, mesh voltage = $-450~\mathrm{V}$, gain $\sim4000$].}
\label{IBF-V}
\end{figure}

The dependence of the ion backflow fraction on the amplification gap \cite{Chefdeville} is shown in Figure \ref{IBF-V-1}.
In each of the cases, the drift field has been kept constant at $200~\mathrm{V/cm}$.
From the figure, it may be concluded that the bulk Micromegas detector with a larger gap has less ion backflow than the bulk Micromegas detector with a smaller gap.
This is due to the fact that for a larger gap, a lower amplification field is required to obtain the same order of gain. At this lower amplification field, $\sigma_{t}$ is larger which ultimately reduces the backflow fraction.
On the other side, due to the larger opening, the backflow fraction is higher for the larger pitch as shown in Figure \ref{IBF-V-2} \cite{Chefdeville}.

In all of the above cases, the simulation follows the experimental trend.
In the simulation, the space charge effect has not been considered which can distort the actual experimental field under certain circumstances.
This may be one of the possible reasons of the discrepancy between the experimental and simulation results, particularly at higher drift field and higher amplification field.
Also, a slight mismatch between the experimentally applied voltage and the simulation voltage can cause the difference between these estimates.

\section{Effect of Double Micro-mesh}

As shown in the earlier section through experimental results and numerical studies, the ion backflow fraction can be controlled by choosing an appropriate geometry and electrostatic configuration.
In this section, we have studied the effect of a second micro-mesh on the electron transmission, the detector gain and, finally, on the ion backflow fraction.

\subsection{Simulation Model}

\begin{figure}[hbt]
\centering
\subfigure[]
{\label{Area-Double}\includegraphics[scale=0.3]{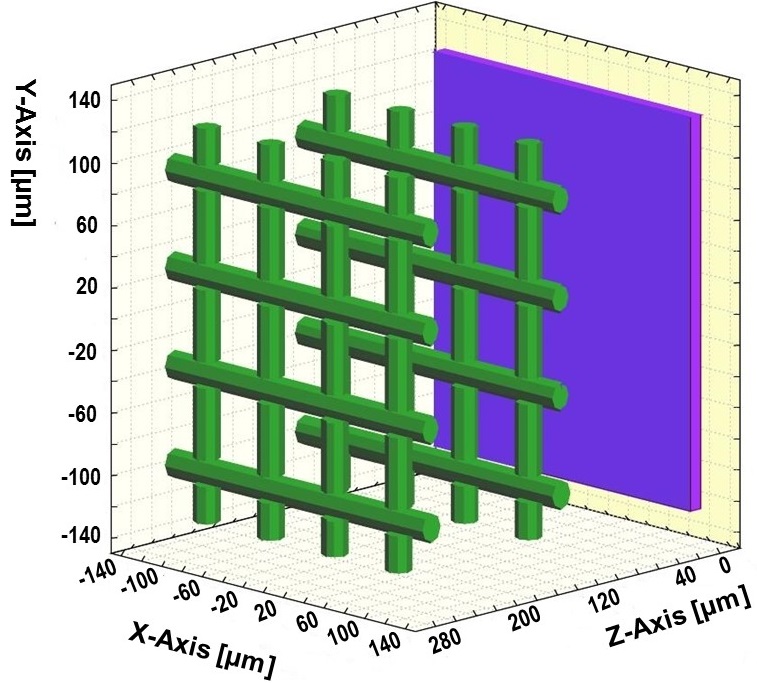}}
\subfigure[]
{\label{Area-Shift0}\includegraphics[scale=0.25]{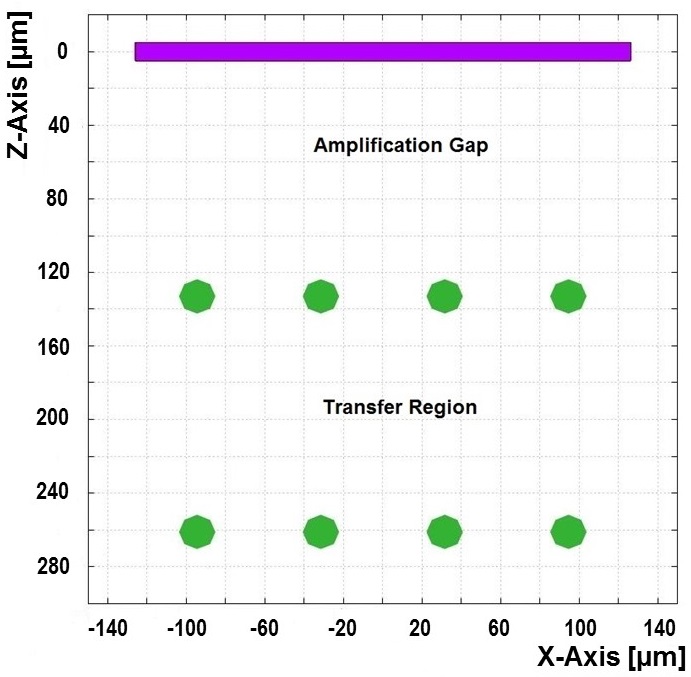}}
\subfigure[]
{\label{Area-Shift16}\includegraphics[scale=0.25]{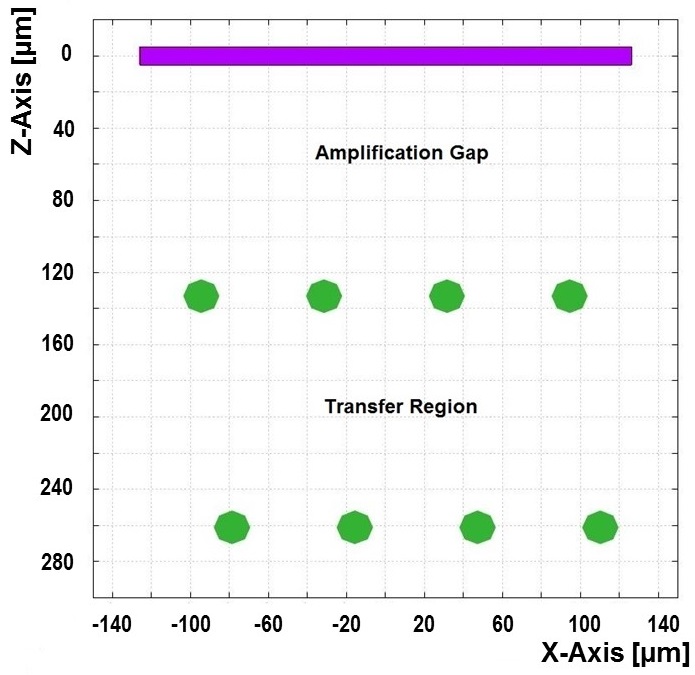}}
\subfigure[]
{\label{Area-Shift31}\includegraphics[scale=0.25]{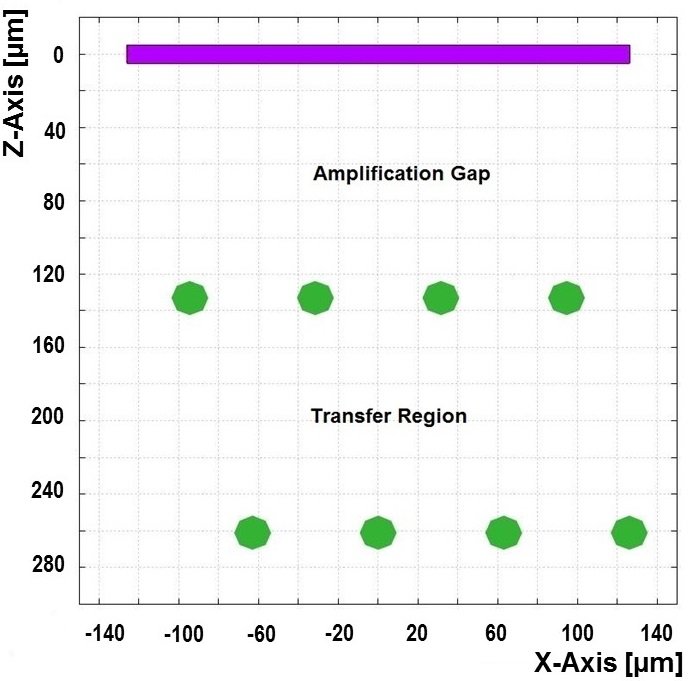}}
\caption{(a) The simulation model for double micro-mesh in 3D; The 2D view, when (b) the two meshes have been placed such that the holes are aligned perfectly and the second micro-mesh is shifted to (c) one quarter ($15.75~\mu\mathrm{m}$) and (d) half the hole pitch ($31.5~\mu\mathrm{m}$) with respect to the first one.}
\label{Area-DoubleMesh}
\end{figure}

As shown in Figure \ref{Area-Double}, a numerical model of a Micromegas having two micro-meshes has been created.
In order to ensure minimum computational complexity, the two micro-meshes have been considered to have the same mesh hole pitch of $63~\mu\mathrm{m}$ (Figure \ref{Area-Double}).
The gap between two micro-meshes, as well as that between the lower micro-mesh and the anode plate, have been assumed to be $128~\mu\mathrm{m}$.
Three different horizontal alignments of the second micro-mesh with respect to the first one have been investigated.
In the first case, the two meshes have been placed such that the holes are aligned perfectly (Figure \ref{Area-Shift0}).
In the other two cases, the second micro-mesh is shifted to one quarter ($15.75~\mu\mathrm{m}$) and half the hole pitch ($31.5~\mu\mathrm{m}$) with respect to the first one.
The alignment schemes are shown in Figure \ref{Area-Shift16} and Figure \ref{Area-Shift31}, respectively.

In this numerical calculation, we have applied the mesh voltages such that the amplification of the electrons occurs in the amplification gap only.
As a result, the upper mesh acts only as a gating mesh and the field in-between the two meshes (denoted as the transfer region) has been adjusted in such a way that it only helps  to transfer the electrons from the drift region to the amplification gap.

\subsection{Electric Field}

\begin{figure}[hbt]
\centering
\includegraphics[scale=0.25]{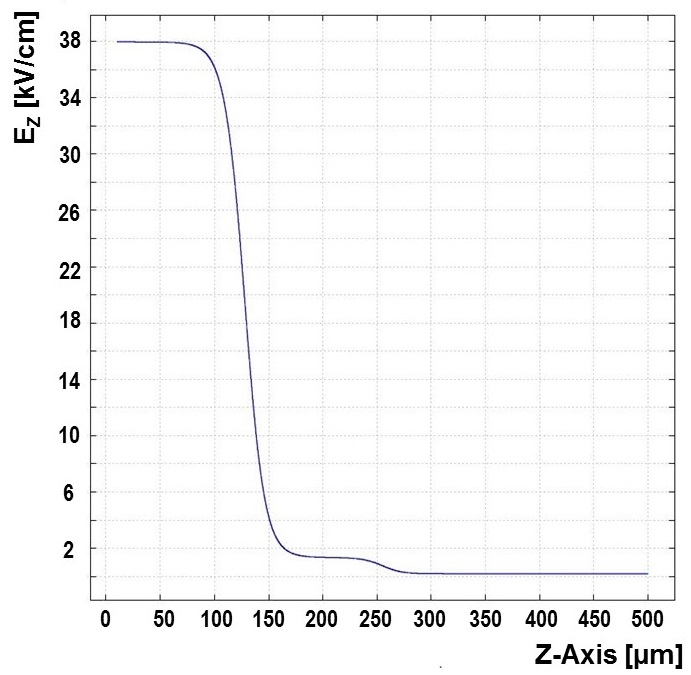}
\caption{The axial electric field through the center of hole for double micro-mesh configuration.}
\label{Field-DoubleMesh}
\end{figure}

The axial electric field through the center of the mesh hole is shown in Figure \ref{Field-DoubleMesh}.
The drift field is $200~\mathrm{V/cm}$ and then in the transfer region, it rises to $1000~\mathrm{V/cm}$ and finally in the amplification gap, it increases further to a value of $38~\mathrm{kV/cm}$.
Depending on the shifted placements, the nature of the field contour changes accordingly.

\subsection{Drift Lines}

\begin{figure}[hbt]
\centering
\subfigure[]
{\label{Drift-Shift0}\includegraphics[scale=0.225]{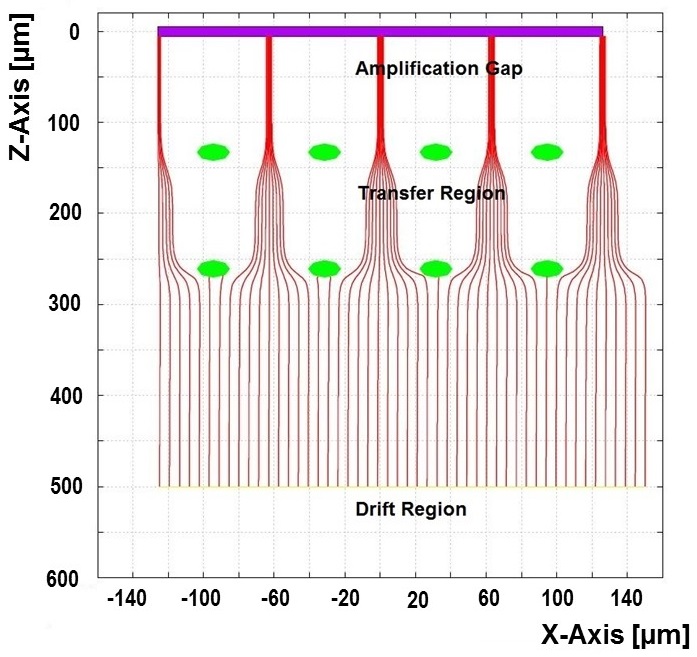}}
\subfigure[]
{\label{Drift-Shift16}\includegraphics[scale=0.225]{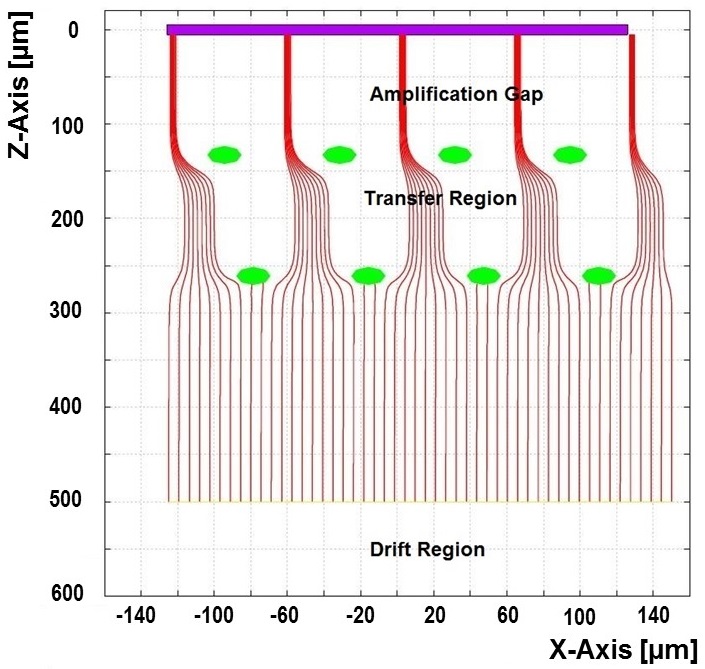}}
\subfigure[]
{\label{Drift-Shift31}\includegraphics[scale=0.225]{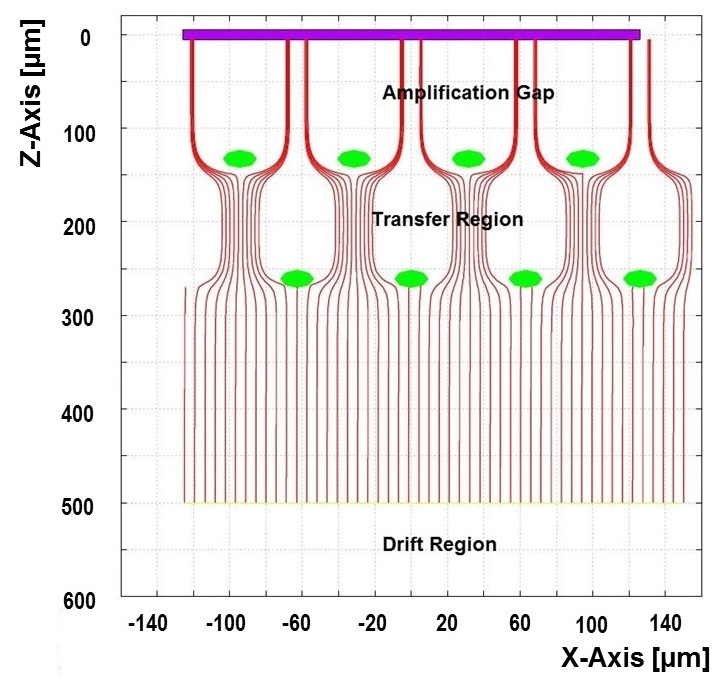}}
\caption{Drift lines for double micro-mesh configurations when the meshes are shifted (a) $0~\mu\mathrm{m}$, (b) $15.75~\mu\mathrm{m}$, (c) $31.5~\mu\mathrm{m}$ with respect to one another.}
\label{Drift-DoubleMesh}
\end{figure}

If diffusion is not considered, the electrons would follow precisely the field lines. Depending on the drift field and the transfer field, the electron drift lines from the drift region are compressed to a funnel shape while entering the transfer region. The field gradient between the transfer field and the amplification field determines the compression factor of the drift lines. The shape of the drift lines is also affected to a great extent by the mutual placement of the holes, as illustrated in figure \ref{Drift-DoubleMesh}. In an actual case, the electron trajectories are also affected significantly by diffusion.

In the following sections, the electron transmission, the gain and the ion backflow have been estimated for the first configuration, {\textit i.e}, when the holes are aligned perfectly. Finally, a comparison with the shifted holes has been preformed. The calculation has been carried out in $\mathrm{Argon}$-$\mathrm{Isobutane}$ (${90:10}$) gas mixture.

\subsection{Electron Transmission}

\begin{figure}[hbt]
\centering
\subfigure[]
{\label{Transmission-100}\includegraphics[scale=0.05]{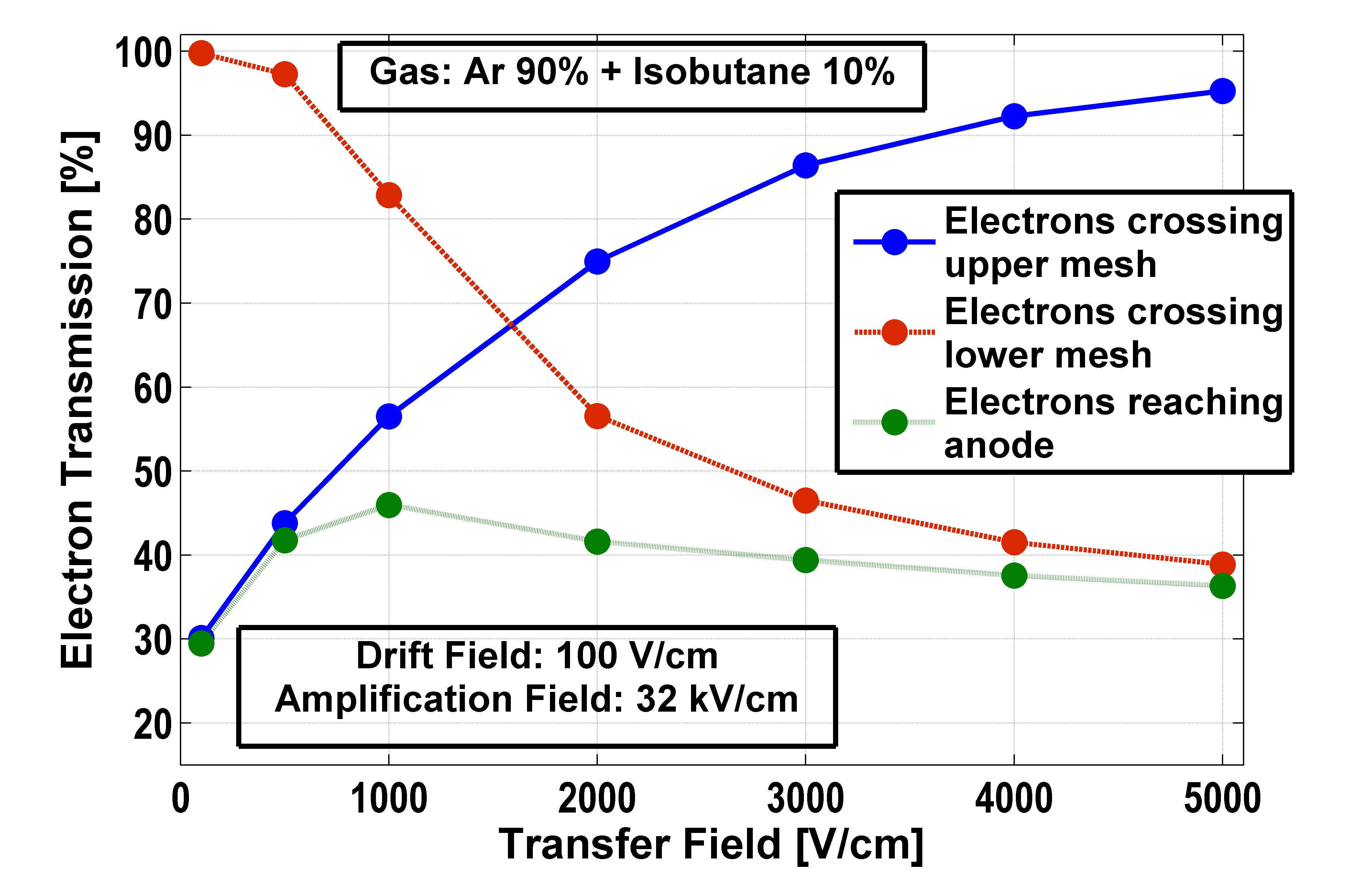}}
\subfigure[]
{\label{Transmission-200}\includegraphics[scale=0.05]{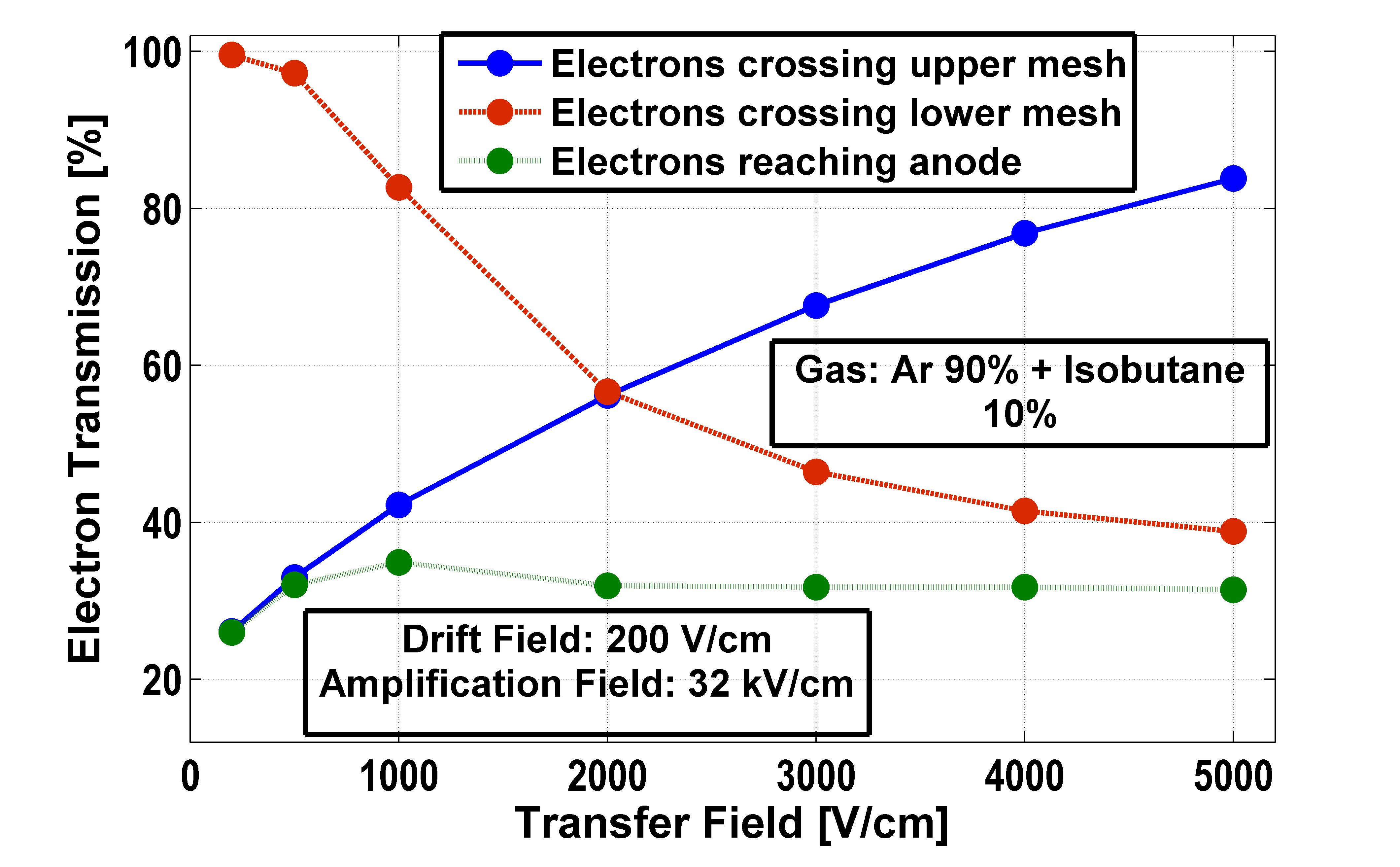}}
\subfigure[]
{\label{Transmission-Amplification}\includegraphics[scale=0.05]{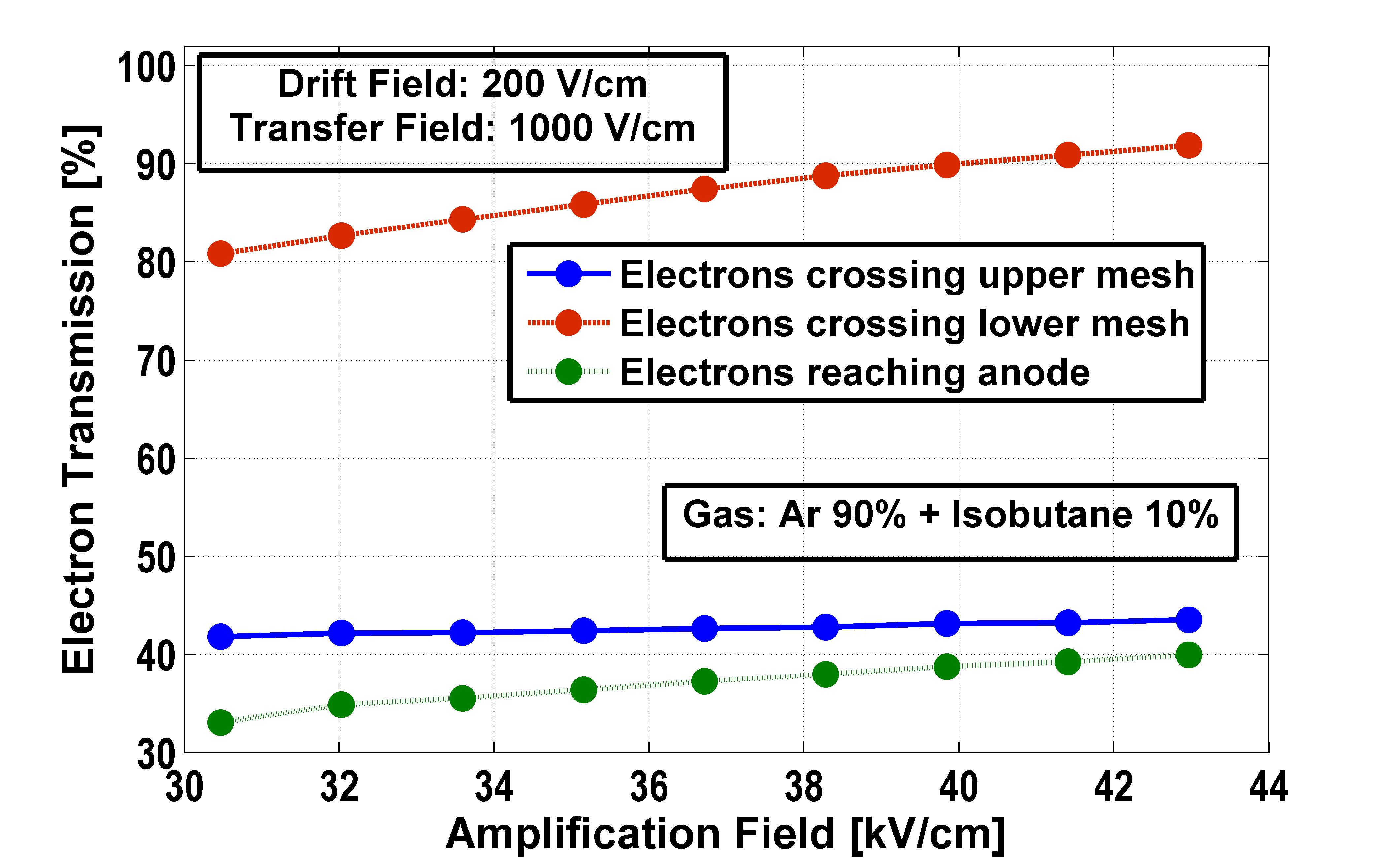}}
\caption{For double micro-mesh, in $\mathrm{Argon}$-$\mathrm{Isobutane}$ (${90:10}$) mixture, the variation of electron transmission with transfer field for drift field of (a) $100~\mathrm{V/cm}$, (b) $200~\mathrm{V/cm}$ at a fixed mesh voltage of $-410~\mathrm{V}$. (c) The same variation but with amplification field at a fixed drift field of $200~\mathrm{V/cm}$ and transfer field of $1000~\mathrm{V/cm}$.}
\label{Transmission-DoubleMesh}
\end{figure}

For a given gas mixture and detector geometry, the electron transmission depends on the field ratios between the drift, transfer and amplification fields.
Microscopic tracking of the electrons has been used in the estimation of the electron transmission in the double micro-mesh device.
The variation of the electron transmission with the transfer field for two different drift fields but the same amplification field is shown in Figure \ref{Transmission-Amplification}.
For a fixed drift field, increasing the transfer field increases the field ratio between the transfer region and the drift volume and, thus, enhances the fraction of electrons which crosses the upper micro-mesh.
But, for a fixed amplification field, an increase of the transfer field worsens the field ratio between the amplification gap and the transfer region and reduces the fraction of electrons which go from the transfer region to the amplification gap.
The total transmission, {\textit{i.e.}}, the fraction of electrons which reaches the amplification gap from the drift volume through the transfer region, is obtained by the multiplication of the above two fractions.
An optimum condition that maximizes the transmission can be reached by an appropriate tuning of the drift and the transfer field. The increase in the amplification field keeps improving the possibility of an electron in the transfer region to reach the amplification gap as shown in Figure \ref{Transmission-Amplification}. 
When the magnitudes of the drift and the transfer fields are fixed, increasing the  amplification field can improve the overall transmission. 
But, the total transmission is significantly low in case of the double micro-mesh.
Only $\sim40\%$ of the primary electrons are able to reach the amplification gap from the drift volume.

\subsection{Gain}

\begin{figure}[hbt]
\centering
\includegraphics[scale=0.05]{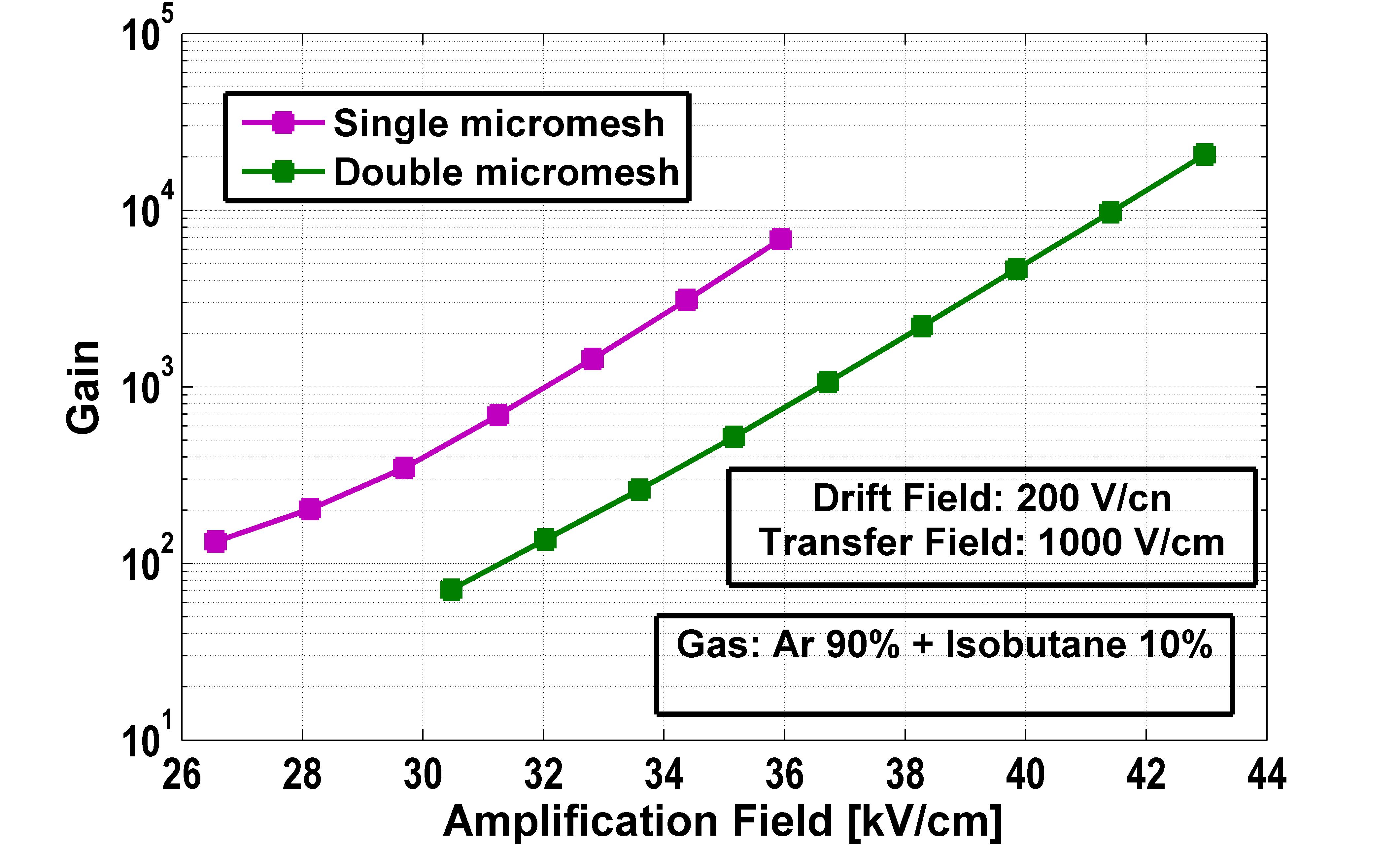}
\caption{Comparison of gain using single and double micro-mesh in $\mathrm{Argon}$-$\mathrm{Isobutane}$ (${90:10}$) mixture.}
\label{Gain-DoubleMesh}
\end{figure}

The use of a double micro-mesh, affects the transmission and, thus, the detector gain.
A comparison with a single mesh configuration is shown in Figure \ref{Gain-DoubleMesh}.
A significantly higher amplification field is necessary to have the same gain for the double mesh device.

\subsection{Ion Backflow}

\begin{figure}[hbt]
\centering
\subfigure[]
{\label{Ion-DoubleMesh-1}\includegraphics[scale=0.3]{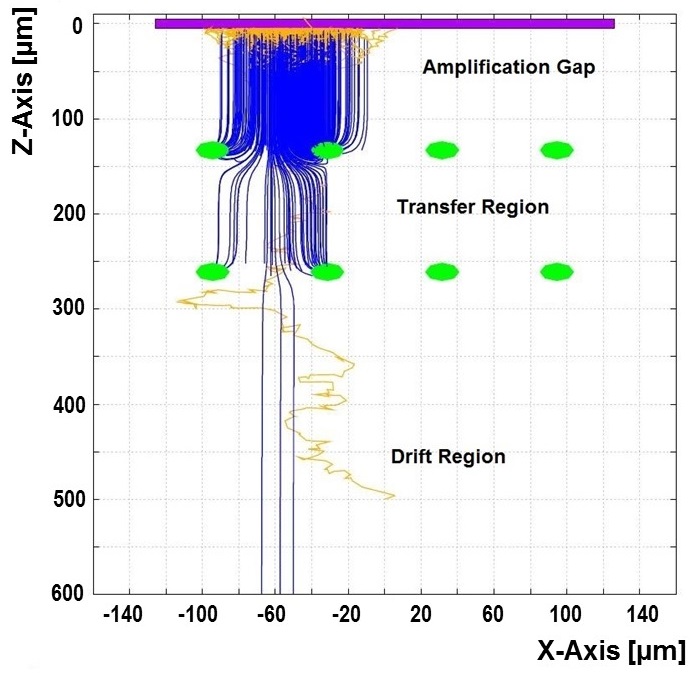}}
\subfigure[]
{\label{Ion-DoubleMesh-2}\includegraphics[scale=0.05]{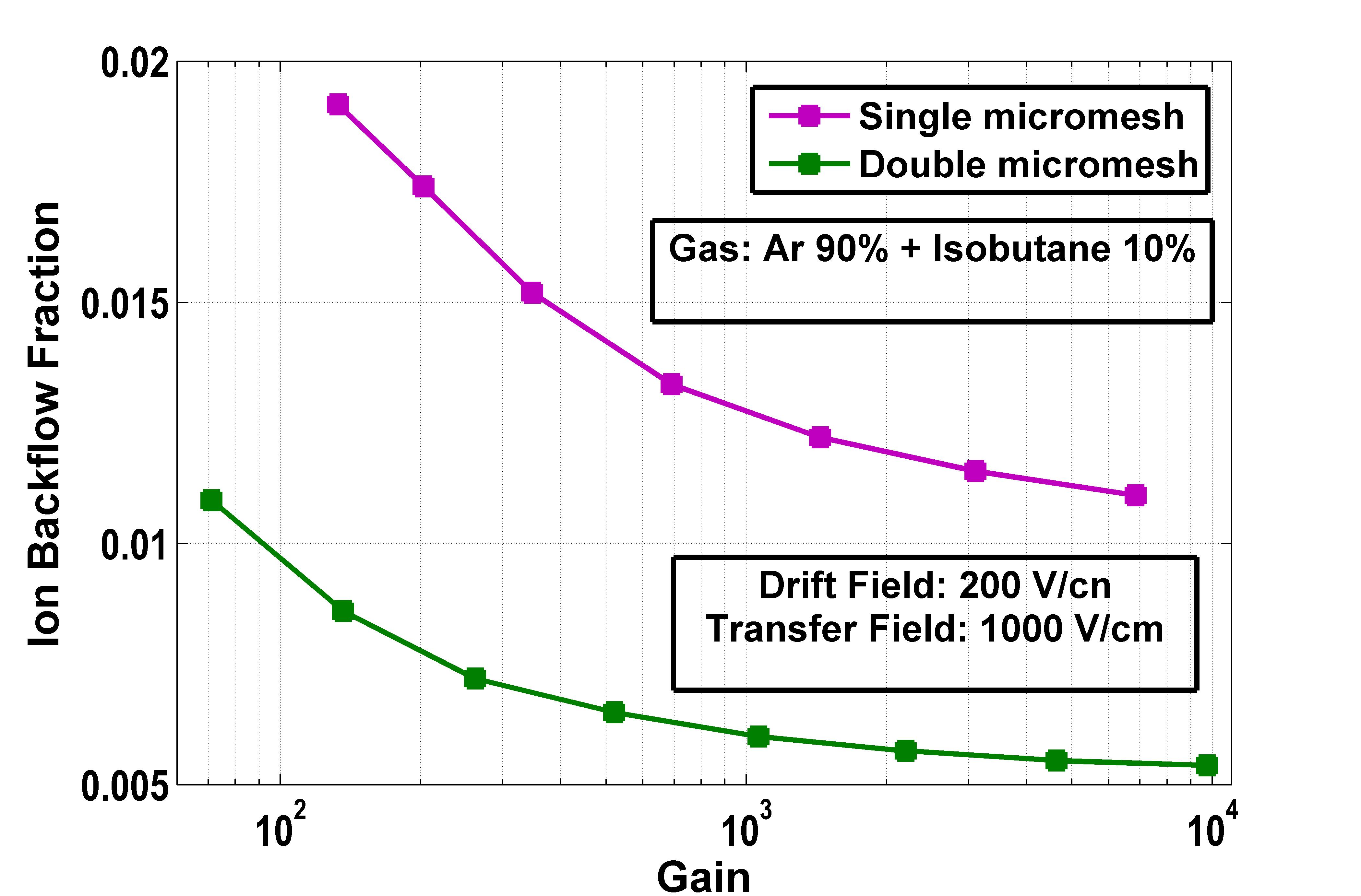}}
\caption{(a) The ion drift lines using double micro-mesh in $\mathrm{Argon}$-$\mathrm{Isobutane}$ (${90:10}$) mixture. (b) the comparison of ion backflow fraction between single and double micro-mesh.}
\label{Ion-DoubleMesh}
\end{figure}

The ion drift lines for a double micro-mesh detector are shown in Figure \ref{Ion-DoubleMesh-1}.
For a particular detector, the ion collection efficiency of the two meshes depends on the field ratios between the drift, transfer and amplification region.
Depending on the field ratios, many of the ions that are not collected by the lower micro-mesh, are collected by the upper micro-mesh.
The variation of the ion backflow fraction with the gain is shown in the same Figure \ref{Ion-DoubleMesh-2}.
A comparison with the single mesh reveals that, at the same value of gain, the use of a double micro-mesh reduces the backflow fraction by a factor of $\sim2$, in qualitative agreement with \cite{IBF2}.

\subsection{Comparison between Three Configuration}

A comparison of the aforesaid characteristics for three different placements of the two micro-meshes has been presented in the Table \ref{Table-DoubleMesh}. It may be observed that a shift in the hole position of the two micro-meshes increases the electron transmission and the gain. However, it also leads to higher ion backflow fractions.

\begin{table}[hbt]
\caption{Electron transmission, gain and ion backflow for three different double mesh configurations in $\mathrm{Argon}$-$\mathrm{Isobutane}$ (${90:10}$) mixture. Amplification gap = $128~\mu\mathrm{m}$, pitch = $63~\mu\mathrm{m}$, transfer region = $128~\mu\mathrm{m}$, mesh voltage = $-470~\mathrm{V}$, drift field = $200~\mathrm{V/cm}$, transfer field = $1000~\mathrm{V/cm}$.}\label{Table-DoubleMesh}
\begin{center}
\begin{tabular}{|c|c|c|c|}
\hline
Shift & Total Electron Transmission & Gain & Ion Backflow Fraction\\
\hline
$0~\mu\mathrm{m}$ & $37.28\%$ & $1063$ & $0.0060$ \\
\hline
$15.75~\mu\mathrm{m}$ & $38.95\%$ & $1140$ & $0.0077$ \\
\hline
$31.5~\mu\mathrm{m}$ & $39.29\%$ & $1207$ & $0.0082$ \\
\hline
\end{tabular}
\end{center}
\end{table}

\section{Conclusions}

In this paper, we have presented the results of experimental and numerical studies illustrating the effects of different geometrical configurations on the ion backflow in bulk Micromegas detector using $\mathrm{Argon}$-$\mathrm{Isobutane}$ ($90:10$) gas mixtures.
Ion backflow fraction has been experimentally estimated using a new experimental setup that employs $^{55}{\mathrm{Fe}}$ as the source of radiation.
The current in different electrodes have been measured by pico-ammeter.
The minimum backflow fraction occuring at  lower drift fields and higher amplification fields confirms our understanding of the earlier results in which a significantly stronger source of radiation was used.
Additional study on the effects of various geometrical parameters of the bulk Micromegas has shown that the detectors with larger gap and smaller pitch has less ion backflow fraction at the stable operating regime.
Comparison of the measured data with the simulation results indicates that the physics of the amplifying micro-mesh-based structures (Micromegas) is understood and can be suitably modeled mathematically.
Finally numerical calculation has been performed using two micro-meshes to explore the effect of this modified geometry on the backflow fraction.
In this context, the influence of this second micro-mesh on the electron transmission and the detector gain has been also studied.
From the above studies, we conclude that the use of a double micro-mesh reduces the backflow fraction by a factor of $\sim2$ compared to that of a single micro-mesh structure.
But this double micro-mesh configuration is not without its problems.
It affects the electron transmission and the gain adversely and, thus, degrades the energy resolution.
Though a shift between the hole positions of the second micro-mesh improves the electron transmission and the gain, the ion backflow also increases.

\section{Acknowledgment}

This work has partly been performed in the framework of the RD51 Collaboration.
We wish to acknowledge the members of the RD51 Collaboration for their help and suggestions.
We are thankful to Abhik Jash for his help in some measurements and Pradipta Kumar Das, Amal Ghoshal for their technical help.
We thank our collaborators from the ILC-TPC collaboration for their help and suggestions.
We also acknowledge Rui de Oliveira and the CERN MPGD workshop for technical support.
We gratefully acknowledge IFCPAR/ CEFIPRA (Project No. 4304-1) for partial financial support. 
Finally, we thank our respective Institutions for providing us with the necessary facilities.

\end{document}